\documentclass[a4paper,12pt,english]{iopart}
\usepackage[T1]{fontenc}
\usepackage[utf8]{inputenc}

\usepackage[usenames,dvipsnames]{xcolor}
\usepackage{graphicx}
\usepackage{iopams}
\usepackage{setstack}
\usepackage{multirow}
\usepackage{siunitx}

\usepackage{cite}

\usepackage{babel}

\newcommand{\SiV}{\mathrm{SiV}^-}

\begin{document}

\title{Isotopically varying spectral features of silicon vacancy in diamond}

\author{
Andreas Dietrich$^1$,
Kay D Jahnke$^1$,
Jan M Binder$^1$,
Tokuyuki Teraji$^2$,
Junichi Isoya$^3$,
Lachlan J Rogers$^1$,
Fedor Jelezko$^1$,
}

\address{$^1$ Institut f\"ur Quantenoptik and IQST, Universit\"at Ulm, D-89081 Ulm, Germany}
\address{$^2$ National Institute for Materials Science, 1-1 Namiki, Tsukuba,
Ibaraki 305-0044, Japan}
\address{$^3$ Research Center for Knowledge Communities, University of Tsukuba,
1-2 Kasuga, Tsukuba, Ibaraki 305-8550, Japan}
\ead{lachlan.j.rogers@quantum.diamonds}

\begin{abstract}
The silicon-vacancy centre ($\SiV$) in diamond has interesting vibronic features. 
We demonstrate that the zero phonon line position can be used to reliably identify the silicon isotope present in a single centre.
This is of interest for quantum information applications since only the $^{29}$Si isotope has nuclear spin.
In addition, we demonstrate that the 64\,meV line is due to a local vibrational
mode of the silicon atom.
The presence of a local mode suggests a plausible origin of the isotopic shift of the zero phonon line.

\end{abstract}

\pacs{}

\maketitle

\section{Introduction}

\begin{figure*}
\includegraphics[width=\textwidth]{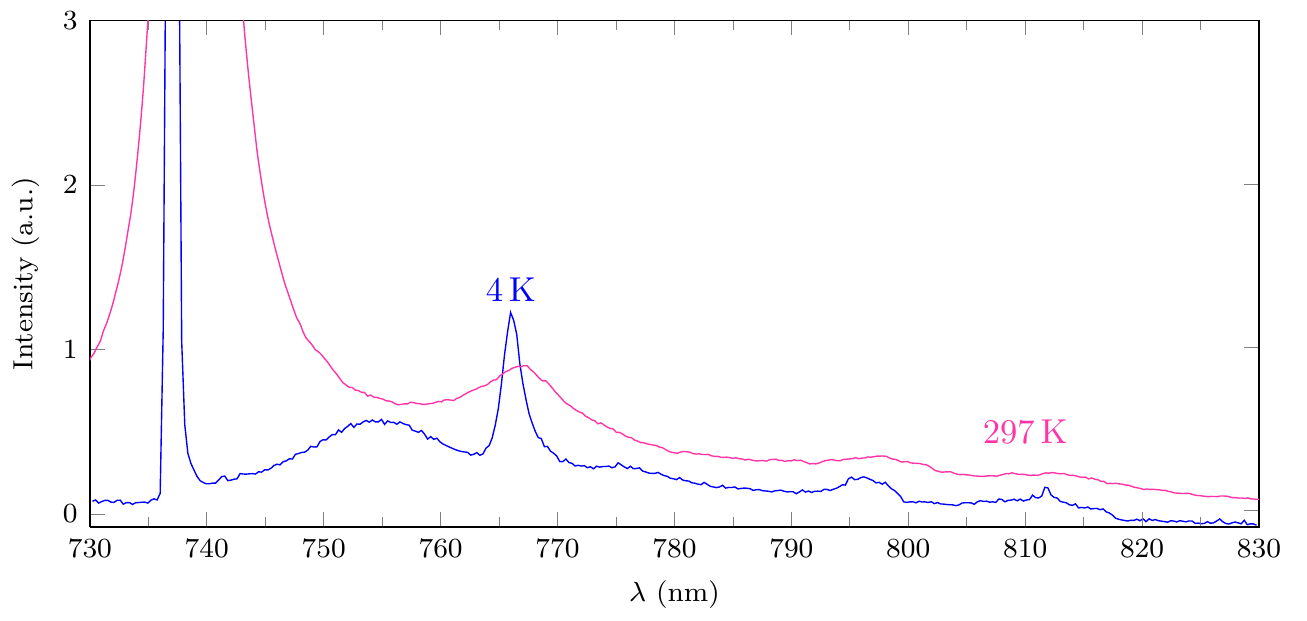}
\caption{
	Typical photoluminescence spectra from a single $\SiV$ site at room
	and cryogenic temperatures. 
	The strong ZPL at 737\,nm contains 70\% (65\%) of the emission at room (cryogenic) temperature.
	The sideband shows a number of features, including a prominent narrow line at about  766 nm which is 64\,meV from the ZPL.
}
\label{fig:spectra}
\end{figure*}

Colour centres in diamond provide attractive architectures for quantum information science and quantum metrology.
They can be detected at the single-site level \cite{gruber_scanning_1997,wang2006single}, and therefore are candidates for single photon generation \cite{kurtsiefer_stable_2000, rogers2013multiple}, quantum information processing \cite{knill_scheme_2001, wrachtrup2006processing, childress2006fault-tolerant}, nano-scale sensing \cite{maletinsky_robust_2012,ermakova2013detection} and bio-marking \cite{fu_characterization_2007, vlasovmolecular-sized2013}.
The negative silicon-vacancy ($\SiV$) centre in diamond has shown potential as an exceptional source for single photon applications \cite{vlasov2009nanodiamond, neu2011single}, and recently multiple $\SiV$ centres have been demonstrated to produce indistinguishable photons intrinsically \cite{rogers2013multiple, sipahigil_indistinguishable_2014}.
It has a strong zero phonon line (ZPL) which contains about 70 \% of the total fluorescence, and a typical photoluminescence spectrum is shown in Figure \ref{fig:spectra}.
Recent steps towards optical access of the $\SiV$ spin have raised the exiting possibility of a usable qubit system being identified in this colour centre \cite{muller2014optical}.
Studies of the related nitrogen vacancy (NV$^-$) centre in diamond have demonstrated that optical control of electronic spin can provide access to nuclear spins of ${}^{13}$C atoms in the lattice \cite{childress_coherent_2006} and ${}^{15}$N atoms forming the defect site \cite{jacques_dynamic_2009}.  
Nuclear spins have superior coherence properties and are ideal for quantum information processing applications \cite{dutt_quantum_2007}.
The most abundant isotope of silicon (${}^{28}$Si) has no nuclear spin, but ${}^{29}$Si is known to have a spin $I=1/2$.
Here we show it is possible to spectrally determine the silicon isotope present in a single $\SiV$ centre, providing an important technique to find nuclear spins.

Progress towards these exciting spin applications also depends on a substantial understanding of the $\SiV$ centre.
Despite a number of recent advancements \cite{gali2013abinitio, neu2013low-temperature, hepp2014electronic, rogers2014electronic}, many fundamental aspects of the $\SiV$ centre have not been explained.
One of the details that assists theoretical modeling is the vibrational behaviour \cite{alkauskas2014first-principles}.  
Although the $\SiV$ sideband is weak, recent polarisation results have suggested interesting physics is displayed in the phonon peaks \cite{rogers2014electronic}.
In this work we examine the sideband for each of the three stable silicon isotopes, and unambiguously confirm that a sharp vibrational peak at 64\,meV is a local mode involving axial oscillation of the silicon atom.
This result resolves contention about the nature of this sideband feature.
The neutral charge state SiV$^{0}$ was established by magnetic resonance measurements to have a split-vacancy structure giving $\mathrm{D}_\mathrm{3d}$ symmetry \cite{edmonds_electron_2008}, however magnetic resonance has not been observed for $\SiV$.
The same $\mathrm{D}_\mathrm{3d}$ symmetry was proposed for $\SiV$ from {\em ab inito} calculations \cite{goss_twelve-line_1996}, and this model is widely accepted \cite{hepp2014electronic,rogers2014electronic}.
The axial oscillation we report here is consistent with this structure, further justifying its broad acceptance.
In addition, the existence of such a local mode suggests a plausible cause of the ZPL isotopic shift.

\section{Experimental design}

Clark et al. \cite{clark1995silicon} reported twelve lines in the ZPL structure for a $\SiV$ ensemble at low temperature.
This was interpreted as a four-line pattern repeated three times corresponding to the three stable isotopes of silicon.
This isotopic shift suggests that it is possible to identify which silicon isotope is present in an individual $\SiV$ centre.
However, many of the early single-site measurements of $\SiV$ were performed in nanodiamonds and showed large site-to-site variation in the $\SiV$ spectrum \cite{neu2013low-temperature}.
This variation makes it difficult to be sure that spectral shifts of single sites are due to the silicon isotope.
Recently, much more uniform spectral lines were reported for single $\SiV$ sites in high-purity bulk diamond \cite{rogers2013multiple}.
That same diamond sample is used here.

The $\SiV$ centres were incorporated into a microwave-plasma chemical vapour deposition (CVD) layer during growth.
The single crystal CVD layer was grown on the \{001\} surface of a low-strain high-pressure high-temperature (HPHT) diamond substrate.
Silicon was introduced into the plasma as it etched 6H-SiC placed on the sample mount, and this process assured a natural abundance of silicon isotopes.
The silicon incorporated during diamond growth produced $\SiV$ sites at a density of about 1.5 sites per \SI{}{\cubic \micro \metre} at the measurement depth of 2--\SI{3}{\micro \metre}.
This density is low enough to resolve single sites in the confocal microscope, but still is high enough to make it convenient to examine a large number of $\SiV$ sites.

The diamond sample was mounted on the cold finger of a continuous-flow helium cryostat, and cooled to about 8\,K.
The MPCVD layer was imaged using a home-built confocal microscope, which had a scan range of 200$\times$\SI{200}{\micro \metre}.
The objective had NA=0.95 and a magnification of 100x.
Excitation was provided by 532\,nm CW laser with about 1\,mW of power at the objective.
Fluorescence from $\SiV$ was measured using avalanche photo diodes (APD). 
Low resolution spectra covering the entire sideband were measured using a spectrometer with a 150-grooves-per-mm grating.
The ZPL was measured with a 1200-grooves-per-mm grating which gave a 16\,GHz resolution capable of resolving the four fine-structure lines separated by 50\,GHz and 200\,GHz.
%

\section{Isotopic shift of spectral features}\label{test}

\begin{figure}
	\includegraphics[width=1\textwidth]{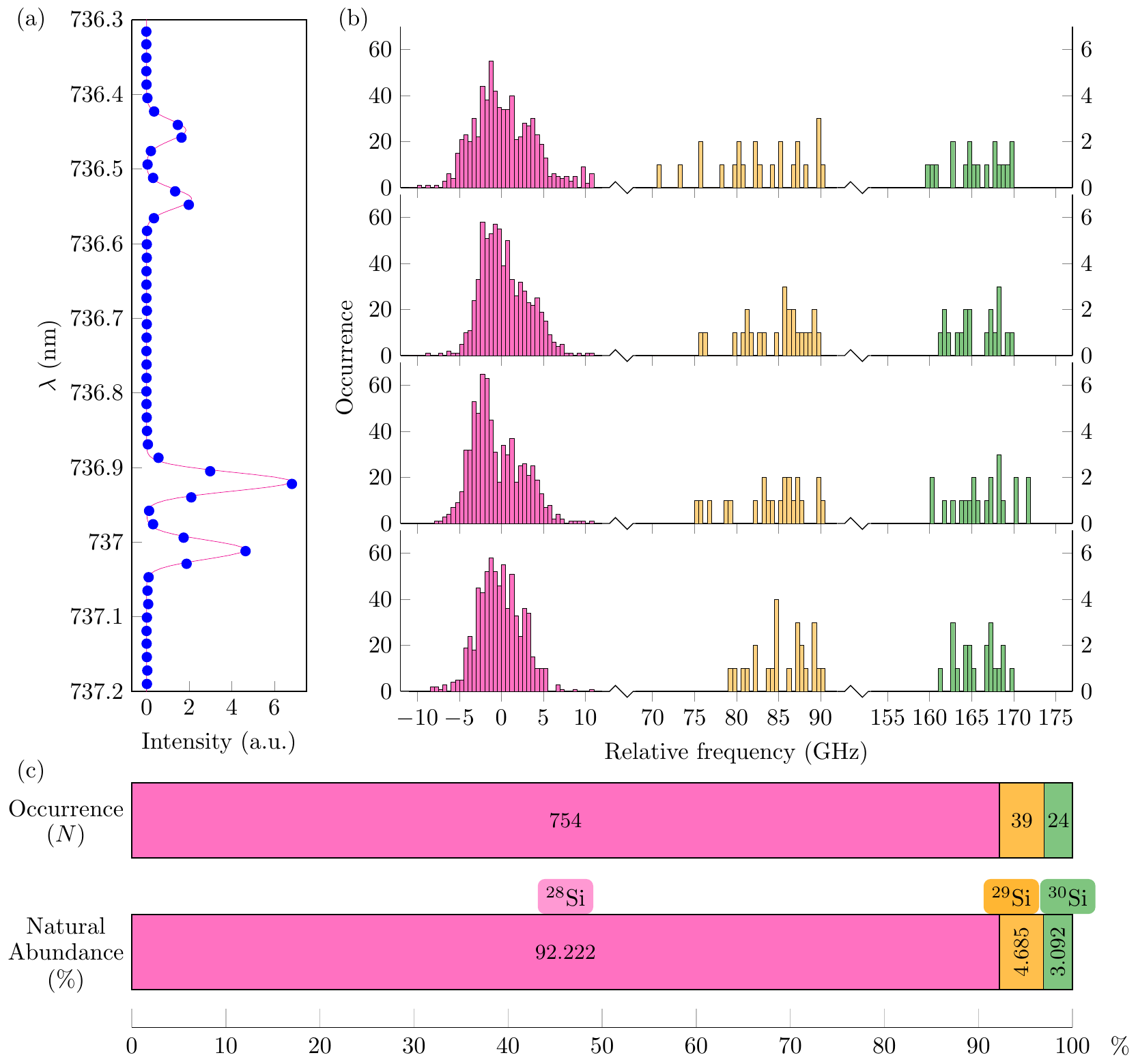}
	\caption{
		Individual $\SiV$ ZPLs form clusters matching the natural abundance of the silicon isotopes. 
		(a) Example spectrum of a typical $\SiV$ fine structure recorded with the spectrometer.
	(b) Histograms showing the distribution of position for each fine-structure line.
	For each line there is strong bunching into three distinct clusters.
	The gap between clusters (about 80\,GHz) is much larger than the width of each cluster (about 12\,GHz) and there were no instances of lines located in between clusters.
	The right-hand vertical axis applies to the centre and right clusters.
	(c) The number of sites in each cluster corresponds to the natural abundance ratios of the three stable silicon isotopes. 
	The occurrence is listed in numbers of measured sites and the abundance in percentage.
		}
	\label{fig:individual_sites}
\end{figure}

Photoluminescence (PL) spectra were recorded for 817 individual $\SiV$ centres at a resolution capable of resolving the ZPL fine structure.
The positions of all four fine-structure lines were extracted from fits to the high-resolution spectra, and binned in intervals of 0.5\,GHz.  
The results are shown in Figure \ref{fig:individual_sites}(b), and it is obvious that all four lines show the $\SiV$ sites naturally clustered into three distinct groups spectrally separated by about 80\,GHz.
The vast majority of these sites were contained in the shortest wavelength cluster, which was found to have an inhomogeneous distribution of about 8\,GHz full-width at half maximum (FWHM).
This is much narrower than the 80\,GHz separation between the clusters, and so the clustering must arise from a physical effect and cannot be due to noise.
The two clusters displaced to longer wavelength occurred less frequently.
The total number of sites in each of these three clusters corresponds closely to the natural abundance of silicon isotopes as illustrated in Figure \ref{fig:individual_sites}(c).
It is concluded that the displacement between clusters must arise from changing the silicon isotope present in the $\SiV$ centre.
This is the first observation of $\SiV$ isotopic shift at the single-site level, and indicates that the silicon isotope present in an individual $\SiV$ site can be unambiguously identified.
The ``digital shift'' between distinct $\SiV$ sites is a elegant confirmation of the isotopic explanation presented for the 12-line structure seen in ensembles \cite{clark1995silicon}.

\begin{figure}
\includegraphics[width=\textwidth]{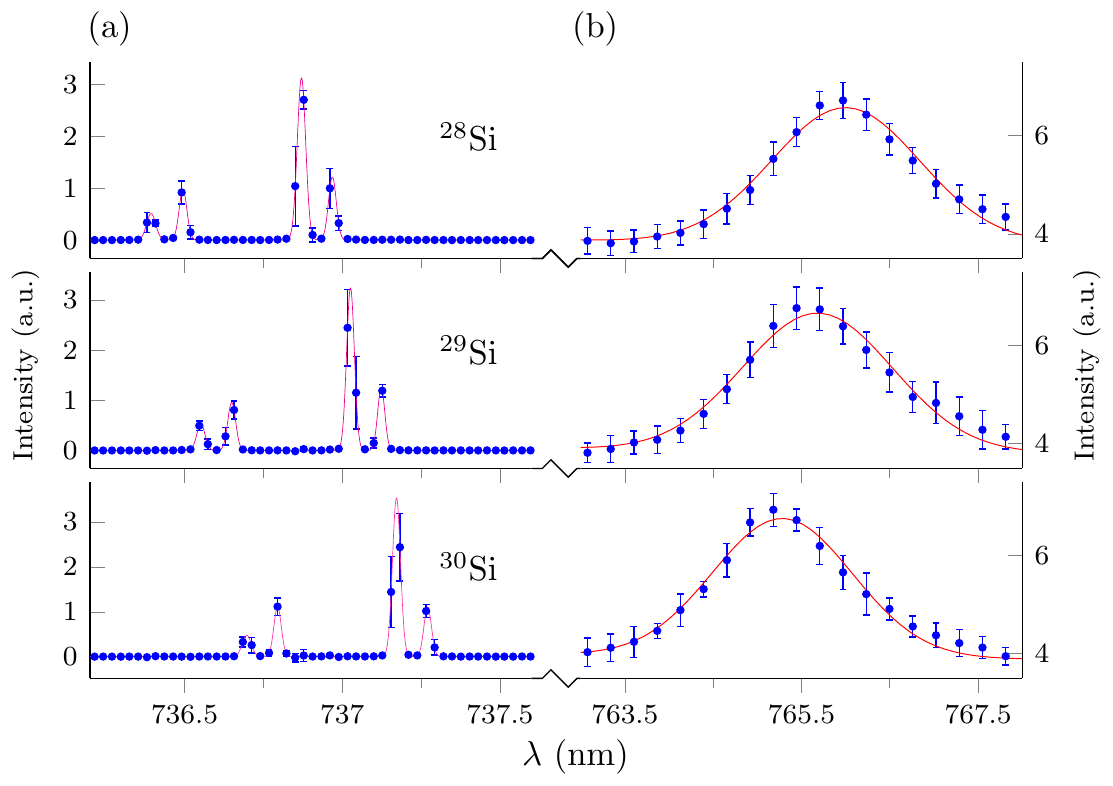}
\caption{
	Two features in the $\SiV$ photolumiescence band shift depending on silicon isotope.
	(a) The ZPL shifts to longer wavelength as the silicon atom increases in mass.  The shifts are 87\,GHz and 166\,GHz for $^{29}$Si and $^{30}$Si, respectively.
	(b) The sharp phonon peak at \SI{64}{\meV} also shifts with the silicon isotope. This feature moves to shorter wavelengths as the silicon mass increases, but the actual energy of the phonon mode is given by the separation between this peak and the ZPL.  Therefore the phonon energy decreases for $^{29}$Si and $^{30}$Si. The fine structure of the ZPL was recorded with 1200 grooves/mm,  and the sideband with 150 grooves/mm.
	The calculated phonon energies for this feature are $63.76 \pm 0.06$\,meV for ${}^{28}$Si and $62.74 \pm 0.06$\,meV for ${}^{29}$Si and $61.55 \pm 0.06$\,meV for ${}^{30}$Si.
The intensities of the ZPL and the sideband spectra cannot be compared since the acquisition was done with different gratings.
}
\label{fig:local_mode}
\end{figure}

The average ZPL spectrum for each silicon isotope is shown in Figure \ref{fig:local_mode}(a).
The fine structure splittings for each isotope are identical, but the ${}^{29}$Si and  ${}^{30}$Si
ZPLs are shifted to longer wavelength by 87\,GHz and 166\,GHz respectively.
These isotopic shifts of the ZPL for $^{29}$Si and $^{30}$Si are in close correspondence with the values previously observed in ensemble measurements \cite{clark1995silicon}.

With this ability to identify the silicon isotope we examine the entire phonon sideband.
The only other feature in the PL spectrum that exhibited measurable isotopic shift was the 64\,meV phonon peak.
The results of the 64\,meV peak for different isotopes are shown in Figure \ref{fig:local_mode}(b).
This feature moves to shorter wavelength meaning that the phonon energy, which is determined by the separation from the ZPL, is decreasing with heavier isotopes.

\section{Identification of local vibrational mode}

It is possible to calculate the precise energy of this phonon peak for each of the silicon isotopes by determining its distance from the corresponding ZPL. 
The phonon energies give the ratios
\begin{equation}
\frac{E_{28}}{E_{29}}=1.0163\pm0.0014 \quad \approx \quad 1.0177 = \sqrt{\frac{m_{29}}{m_{28}} }
\label{64mevRatio29}
\end{equation} 
and
\begin{equation}
\frac{E_{28}}{E_{30}}=1.0359\pm0.0011 \quad \approx \quad 1.0357 = \sqrt{\frac{m_{30}}{m_{28}} }
\label{64mevRatio30}
\end{equation} 
These ratios are in close agreement with a simple harmonic oscillator model where the phonon frequency $\omega$ is given by
\begin{equation}
\omega=\sqrt{\frac{k}{m}}
\label{omega}
\end{equation} 
for spring constant $k$ and an oscillating mass $m$ corresponding to the silicon atom.
The validity of this simple model indicates this spectral feature arises from a local oscillation of the silicon atom, with weak coupling to the carbon nuclei of the diamond lattice.
Therefore it must be a purely local vibrational mode of the $\SiV$ centre.

This informs an ongoing discussion about this feature in the $\SiV $ sideband, and resolves some contention.
Early in the analysis of the $\SiV$ sideband it was suggested that the peak at 43 meV is due to a local vibrational mode, and that the other features arise from the lattice \cite{feng1993characteristics}.
In contrast, Gorokhovsky \cite{gorokhovsky_photoluminescence_1995} showed that 64\,meV  was close to a phonon energy in silicon and concluded that this $\SiV$ feature was a local mode.  
They also suggested the feature at 128\,meV could be a second harmonic of this vibration. 
More recently it has been proposed that either the 42\,meV or the 64\,meV features could arise from a local mode, depending on the symmetry of the defect \cite{zaitsev_vibronic_2000}.
We have shown conclusively that the 64\,meV sideband feature is due to a local vibrational mode. 
Our spectral measurements did not show any isotopic variation in the 42\,meV feature, although it is significantly broader and so small shifts in position would be difficult to detect.
Additionally, broad peaks do not generally arise from local modes, and therefore we conclude the 42\,meV feature to be non localized.
We also observed no isotopic variation in the 128\,meV feature, and conclude that it is not a harmonic of the local mode.

Numerous extra electronic transitions have been proposed to account for the photon autocorrelation statistics of SiV, which exhibit bunching shoulders typical of storage in metastable states  \cite{neu2012photophysics}.
An additional electronic transition has been reported at 822.7\,nm \cite{neu_electronic_2012}, although this spectral feature was not observed in our measurements here.  
Since we have shown that the 64\,meV feature is due to a local vibrational mode, it cannot be due to an electronic transition associated with metastable states.  
We conclude that the 64\,meV sideband peak does not give insight to the storage mechanism.

\begin{figure}
	\centering
		\includegraphics[width=5cm]{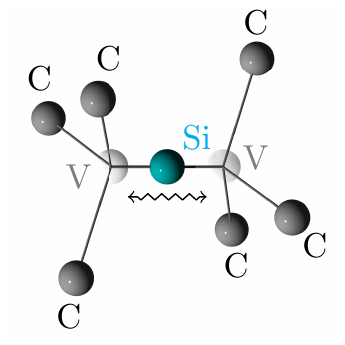}
	\caption{
		The $\mathrm{D}_\mathrm{3d}$ symmetry of $\SiV$ with the oscillating silicon atom.
		Carbon (C) atoms are marked in grey, while silicon (Si) is symbolised in cyan.
		The position of the vacancies (V) is marked by opaque spheres.
		This symmetry supports a localised oscillation of the silicon atom along the $\langle 111 \rangle$ axis.
		}
	\label{fig:D3d}
\end{figure}

It is widely accepted that $\SiV$ has $\mathrm{D}_\mathrm{3d}$ symmetry \cite{goss_twelve-line_1996, rogers2014electronic, hepp2014electronic}, which consists of the silicon atom in the middle of a split vacancy as shown in Figure \ref{fig:D3d}.
It is worthwhile considering the implications of a local silicon vibration within this geometry.
The 64\,meV peak is polarized similarly to the ZPL and has most of its strength coming from the axial dipole moment \cite{rogers2014electronic}.
This suggests that the silicon atom oscillates along the $\langle 111 \rangle$ symmetry  axis.  
Since the silicon atom lies between two vacant lattice sites it is weakly bound along this axis. 
Such a geometry is likely to support an oscillation of the silicon atom that does not couple strongly to the carbon lattice.

In this work we have identified the silicon isotope present in each $\SiV$ centre from the ZPL position.
This technique is possible when the inhomogeneous distribution is smaller than the $\approx 80$\,GHz isotopic shift.  
However, in samples with less uniform $\SiV$ sites (such as typically result from Si implantation) it may be possible to use the energy of this local vibrational mode to identify the silicon isotope.

\section{Origin of ZPL isotopic shift}
Although the isotopic shift of the $\SiV$ ZPL has been known since the early ensemble measurements of Clark and Kanda \cite{clark1995silicon}, a mechanism for this shift has never been proposed.
Our observation of a local vibrational mode provides a plausible explanation, following the direction of Iakoubovskii and Davies \cite{iakoubovskii_vibronic_2004} for the 1.4 eV optical center and Lawson et al. with the H2 center \cite{lawson1992h2}.
The key idea is to consider that a local vibrational mode couples to the electronic states of the center.
Hence, the ZPL energy is composed of a purely electronic component and a vibrational component. 
This additional energy depends on the curvature of the vibrational potential, which in turn depends on the symmetry properties of the electronic states and therefore differs between ground and excited states. 
For an isotopic shift of the ZPL, this curvature must differ between the ground and excited states.

We have observed a local mode for $\SiV$ which could interact with the ZPL in this manner.
The ZPL energy $h\nu$ in this model can be expressed as a function temperature of the form
\begin{equation}
h \nu(T)= h\nu_\text{el}+\sum_i^N \left( n_i + \frac{1}{2} \right) \hbar \left( \omega^\prime_i-\omega_i \right)
\label{eq:Lawson}
\end{equation}
following Lawson et al.\cite{lawson1992h2}.
Here $h\nu_\text{el}$ is the purely electronic transition energy.
The term in the sum is the energy difference between the excited ($\omega^\prime$) and ground ($\omega$) states of the $i^\mathrm{th}$ phonon mode (which could be a local vibration) at the same occupation level $n$. 
Transitions of this kind between levels of matching phonon occupation contribute to the ZPL since phonons are not involved \cite{iakoubovskii_vibronic_2004}.
This energy is summed over all $N$ phonon modes that couple to electronic states.
The temperature dependence appears only in the vibrational states. 
At zero temperature only the $n=0$ levels will be occupied, while at higher temperatures energy levels with increasing $n$ occur providing a broadening of the ZPL \cite{iakoubovskii_vibronic_2004}.

We do not know which global modes are involved in the sum, but do know that one of the vibrational modes is entirely local.
Because this vibration behaves exactly as a harmonic oscillator, and therefore does not involve the carbon atoms of the diamond lattice, it is possible to separate local modes out from the sum. 
The sum over all phonons can hence be separated into independent sums over the local vibrational modes of the silicon atom and over the phonons of the carbon lattice.
So equation (\ref{eq:Lawson}) becomes
\begin{equation}
h \nu(T)= h\nu_\text{el}+\sideset{}{_\mathrm{Si}}\sum^{M}_i \left( n_i + \frac{1}{2} \right) \hbar \left( \omega^\prime_i-\omega_i \right)+\sideset{}{_\mathrm{C}}\sum^{N-M}_i \left( n_i + \frac{1}{2} \right) \hbar \left( \omega^\prime_i-\omega_i \right).
\label{blalal}
\end{equation}
The Si subscripted sum covers the $M$ local silicon modes, while the
subscript C sums over the $N-M$ carbon phonons.
To see if this model is valid to explain the shift of the ZPL, the energy
ratios between the isotopes are now compared.

As we have established in the previous section, changing the silicon mass only affects the energy of the local phonons which change as $\omega\propto 1/\sqrt{m}$.
Therefore the only changing term is the sum over the local silicon phonons, and so this term gives the differences $\Delta E_{28,29}$ and $\Delta E_{28,30}$.
The frequencies ($\omega_{i}$ and $\omega^\prime_{i}$) of the $^{29}$Si and $^{30}$Si can be expressed in relation to 
${}^{28}$Si via the fraction of the square roots of the masses.
This allows the energy difference to be expressed as:
\begin{equation}
\Delta E_{28,29}=\hbar \left(1-\sqrt{m_{28}/m_{29}}\right)\cdot\left(\sideset{}{_{{}^{28}\mathrm{Si}}}\sum^{M}_{i} \left( n_i + \frac{1}{2} \right) \hbar \left( \omega^\prime_i-\omega_i \right)\right)
\end{equation}
and similarly for $\Delta E_{28,30}$.
The ratio between these energy differences is then
\begin{equation}
\frac{\Delta E_{28,29}}{\Delta E_{28,30}}=\frac{1-\sqrt{m_{28}/m_{29}}}{1-\sqrt{m_{28}/m_{30}}} \approx 0.5135
\end{equation}
Since we have measured the ZPL energy shift for all three silicon isotopes this model can be compared to our experiments.
The calculated ratio is in close agreement with the empirically determined value of $0.52\pm0.07$.

The correspondence between theory and experiment suggests there is validity to this model and that the shift of the ZPL with different isotopes arises from the presence of local modes. 
Without knowing the difference in curvature of the harmonic potentials between the ground and excited states, the model is unable to predict the direction of the ZPL shift.
We have experimentally found that the ZPL moves to lower energy as the silicon mass increases, and this observation may assist theoretical efforts to describe the phonon harmonic potentials.

\section{Conclusion}
We have demonstrated the ability to unambiguously identify the silicon isotope contained in a single $\SiV$ centre from the PL spectra.
This ability is of interest because it allows easy identification of centres containing ${}^{29}$Si.
Only these centres provide the possibility of accessing silicon nuclear spin, which is a feature of interest for quantum information applications.
Nuclear spins in diamond are weakly coupled to their environment and generally have long coherence times which are necessary for quantum information storage.

We identify the 64\,meV phonon sideband feature to be a local vibrational mode of the silicon atom.
This resolves some contention about local modes in the literature \cite{gorokhovsky_photoluminescence_1995,feng1993characteristics,zaitsev_vibronic_2000} and gives insight into the vibrational properties of $\SiV$.
The presence of a local mode suggests a plausible explanation for the ZPL isotopic shift in a similar manner to that proposed for other colour centres in diamond including the 1.4\,eV and H2 centres.
Although the $\mathrm{D}_\mathrm{3d}$ symmetry of $\SiV$ is widely accepted, it has not been directly observed. 
The local vibrational mode we have described here is entirely consistent with $\mathrm{D}_\mathrm{3d}$ symmetry and this result provides further indirect evidence supporting this split-vacancy defect geometry.

\section*{Acknowledgements}
The authors acknowledge funding from EU (DIAMANT, SIQS, DIADEMS), ERC,
German Science Foundation - DFG (SFB TR21, FOR1482, FOR1493), JST, JSPS KAKENHI (No. 26246001), DARPA, Sino-German Center, VW Stiftung.

\nocite{*}
\section*{References}
\bibliographystyle{unsrt}
\bibliography{siv_isotopic_shift}

\end{document}